\documentclass[10pt,a4paper,onecolumn]{article}
\usepackage{marginnote}
\usepackage{graphicx}
\usepackage{xcolor}
\usepackage{authblk,etoolbox}
\usepackage{titlesec}
\usepackage{calc}
\usepackage{tikz}
\usepackage{hyperref}
\hypersetup{colorlinks,breaklinks=true,
            urlcolor=[rgb]{0.0, 0.5, 1.0},
            linkcolor=[rgb]{0.0, 0.5, 1.0}}
\usepackage{caption}
\usepackage{tcolorbox}
\usepackage{amssymb,amsmath}
\usepackage{ifxetex,ifluatex}
\usepackage{seqsplit}
\usepackage{xstring}

\usepackage{float}
\let\origfigure\figure
\let\endorigfigure\endfigure
\renewenvironment{figure}[1][2] {
    \expandafter\origfigure\expandafter[H]
} {
    \endorigfigure
}

\usepackage{fixltx2e} % provides \textsubscript
\usepackage[
  backend=biber,
]{biblatex}
\bibliography{paper.bib}

\let\textttOrig=\texttt
\def\texttt#1{\expandafter\textttOrig{\seqsplit{#1}}}
\renewcommand{\seqinsert}{\ifmmode
  \allowbreak
  \else\penalty6000\hspace{0pt plus 0.02em}\fi}

\makeatletter
\let\href@Orig=\href
\def\href@Urllike#1#2{\href@Orig{#1}{\begingroup
    \def\Url@String{#2}\Url@FormatString
    \endgroup}}
\def\href@Notdoi#1#2{\def\tempa{#1}\def\tempb{#2}%
  \ifx\tempa\tempb\relax\href@Urllike{#1}{#2}\else
  \href@Orig{#1}{#2}\fi}
\def\href#1#2{%
  \IfBeginWith{#1}{https://doi.org}%
  {\href@Urllike{#1}{#2}}{\href@Notdoi{#1}{#2}}}
\makeatother

\newlength{\cslhangindent}
\newlength{\csllabelwidth}
\newenvironment{CSLReferences}[3] % #1 hanging-ident, #2 entry spacing
 {% don't indent paragraphs
  \setlength{\parindent}{0pt}
  \ifodd #1 \everypar{\setlength{\hangindent}{\cslhangindent}}\ignorespaces\fi
  \ifnum #2 > 0
  \setlength{\parskip}{#2\baselineskip}
  \fi
 }%
 {}
\usepackage{calc}

\usepackage[top=3.5cm, bottom=3cm, right=1.5cm, left=1.0cm,
            headheight=2.2cm, reversemp, includemp, marginparwidth=4.5cm]{geometry}

\titleformat{\section}
  {\normalfont\sffamily\Large\bfseries}
  {}{0pt}{}
\titleformat{\subsection}
  {\normalfont\sffamily\large\bfseries}
  {}{0pt}{}
\titleformat{\subsubsection}
  {\normalfont\sffamily\bfseries}
  {}{0pt}{}
\titleformat*{\paragraph}
  {\sffamily\normalsize}

\usepackage{fancyhdr}
\makeatletter
\let\ps@plain\ps@fancy
\fancyheadoffset[L]{4.5cm}
\fancyfootoffset[L]{4.5cm}

\definecolor{linky}{rgb}{0.0, 0.5, 1.0}

\newtcolorbox{repobox}
   {colback=red, colframe=red!75!black,
     boxrule=0.5pt, arc=2pt, left=6pt, right=6pt, top=3pt, bottom=3pt}

\newcommand{\ExternalLink}{%
   \tikz[x=1.2ex, y=1.2ex, baseline=-0.05ex]{%
       \begin{scope}[x=1ex, y=1ex]
           \clip (-0.1,-0.1)
               --++ (-0, 1.2)
               --++ (0.6, 0)
               --++ (0, -0.6)
               --++ (0.6, 0)
               --++ (0, -1);
           \path[draw,
               line width = 0.5,
               rounded corners=0.5]
               (0,0) rectangle (1,1);
       \end{scope}
       \path[draw, line width = 0.5] (0.5, 0.5)
           -- (1, 1);
       \path[draw, line width = 0.5] (0.6, 1)
           -- (1, 1) -- (1, 0.6);
       }
   }

\patchcmd{\@maketitle}{center}{flushleft}{}{}
\patchcmd{\@maketitle}{center}{flushleft}{}{}
\patchcmd{\@maketitle}{\LARGE}{\LARGE\sffamily}{}{}
\def\maketitle{{%
  
  \AB@maketitle}}
\makeatletter
\renewcommand\AB@affilsepx{ \protect\Affilfont}
\renewcommand\AB@affilnote[1]{{\bfseries #1}\hspace{3pt}}
\renewcommand{\affil}[2][]%
   {\newaffiltrue\let\AB@blk@and\AB@pand
      \if\relax#1\relax\def\AB@note{\AB@thenote}\else\def\AB@note{#1}%
        \setcounter{Maxaffil}{0}\fi
        \begingroup
        \let\href=\href@Orig
        \let\texttt=\textttOrig
        \let\protect\@unexpandable@protect
        \def\thanks{\protect\thanks}\def\footnote{\protect\footnote}%
        \@temptokena=\expandafter{\AB@authors}%
        {\def\\{\protect\\\protect\Affilfont}\xdef\AB@temp{#2}}%
         \xdef\AB@authors{\the\@temptokena\AB@las\AB@au@str
         \protect\\[\affilsep]\protect\Affilfont\AB@temp}%
         \gdef\AB@las{}\gdef\AB@au@str{}%
        {\def\\{, \ignorespaces}\xdef\AB@temp{#2}}%
        \@temptokena=\expandafter{\AB@affillist}%
        \xdef\AB@affillist{\the\@temptokena \AB@affilsep
          \AB@affilnote{\AB@note}\protect\Affilfont\AB@temp}%
      \endgroup
       \let\AB@affilsep\AB@affilsepx
}
\makeatother

\renewcommand\Affilfont{\sffamily\small\mdseries}
\ifnum 0\ifxetex 1\fi\ifluatex 1\fi=0 % if pdftex
  \usepackage[T1]{fontenc}
  \usepackage[utf8]{inputenc}

\else % if luatex or xelatex
  \ifxetex
    \usepackage{mathspec}
    \usepackage{fontspec}

  \else
    \usepackage{fontspec}
  \fi
  \defaultfontfeatures{Ligatures=TeX,Scale=MatchLowercase}

\fi
\IfFileExists{upquote.sty}{\usepackage{upquote}}{}
\IfFileExists{microtype.sty}{%
\usepackage{microtype}
\UseMicrotypeSet[protrusion]{basicmath} % disable protrusion for tt fonts
}{}

\usepackage{hyperref}
\hypersetup{unicode=true,
            pdftitle={Brahe: A Modern Astrodynamics Library for Research and Engineering Applications},
            pdfborder={0 0 0},
            breaklinks=true}
\let\addcontentslineOrig=\addcontentsline
\def\addcontentsline#1#2#3{\bgroup
  \let\texttt=\textttOrig\addcontentslineOrig{#1}{#2}{#3}\egroup}
\let\markbothOrig\markboth
\def\markboth#1#2{\bgroup
  \let\texttt=\textttOrig\markbothOrig{#1}{#2}\egroup}
\let\markrightOrig\markright
\def\markright#1{\bgroup
  \let\texttt=\textttOrig\markrightOrig{#1}\egroup}

\usepackage{graphicx,grffile}
\makeatletter
\def\maxwidth{\ifdim\Gin@nat@width>\linewidth\linewidth\else\Gin@nat@width\fi}
\def\maxheight{\ifdim\Gin@nat@height>\textheight\textheight\else\Gin@nat@height\fi}
\makeatother
\setkeys{Gin}{width=\maxwidth,height=\maxheight,keepaspectratio}
\IfFileExists{parskip.sty}{%
\usepackage{parskip}
}{% else
\setlength{\parindent}{0pt}
\setlength{\parskip}{6pt plus 2pt minus 1pt}
}
\ifx\paragraph\undefined\else
\let\oldparagraph\paragraph
\renewcommand{\paragraph}[1]{\oldparagraph{#1}\mbox{}}
\fi
\ifx\subparagraph\undefined\else
\let\oldsubparagraph\subparagraph
\renewcommand{\subparagraph}[1]{\oldsubparagraph{#1}\mbox{}}
\fi
\usepackage{listings}

\title{Brahe: A Modern Astrodynamics Library for Research and
Engineering Applications}

        \author[1]{Duncan Eddy}
          \author[1]{Mykel J. Kochenderfer}
    
      \affil[1]{Stanford University}
  \date{\vspace{-7ex}}

\begin{document}
\maketitle

\marginpar{

  \begin{flushleft}
  %\hrule
  \sffamily\small

  {\bfseries DOI:} \href{https://doi.org/DOI unavailable}{\color{linky}{DOI unavailable}}

  \vspace{2mm}

  {\bfseries Software}
  \begin{itemize}
    \setlength\itemsep{0em}
    \item \href{N/A}{\color{linky}{Review}} \ExternalLink
    \item \href{NO_REPOSITORY}{\color{linky}{Repository}} \ExternalLink
    \item \href{DOI unavailable}{\color{linky}{Archive}} \ExternalLink
  \end{itemize}

  \vspace{2mm}

  \par\noindent\hrulefill\par

  \vspace{2mm}

  {\bfseries Editor:} \href{https://example.com}{Pending
Editor} \ExternalLink \\
  \vspace{1mm}
    {\bfseries Reviewers:}
  \begin{itemize}
  \setlength\itemsep{0em}
    \item \href{https://github.com/Pending Reviewers}{@Pending
Reviewers}
    \end{itemize}
    \vspace{2mm}

  {\bfseries Submitted:} N/A\\
  {\bfseries Published:} N/A

  \vspace{2mm}
  {\bfseries License}\\
  Authors of papers retain copyright and release the work under a Creative Commons Attribution 4.0 International License (\href{http://creativecommons.org/licenses/by/4.0/}{\color{linky}{CC BY 4.0}}).

  \end{flushleft}
}

\lstset{
    language         = Python,
    backgroundcolor  = \color[HTML]{F2F2F2},
    basicstyle       = \small\ttfamily\color[HTML]{19177C},
    numberstyle      = \ttfamily\scriptsize\color[HTML]{7F7F7F},
    keywordstyle     = [1]{\bfseries\color[HTML]{1BA1EA}},
    keywordstyle     = [2]{\color[HTML]{0F6FA3}},
    keywordstyle     = [3]{\color[HTML]{0000FF}},
    stringstyle      = \color[HTML]{F5615C},
    commentstyle     = \color[HTML]{AAAAAA},
    rulecolor        = \color[HTML]{000000},
    frame=lines,
    xleftmargin=10pt,
    framexleftmargin=10pt,
    framextopmargin=4pt,
    framexbottommargin=4pt,
    tabsize=4,
    captionpos=b,
    breaklines=true,
    breakatwhitespace=false,
    showstringspaces=false,
    showspaces=false,
    showtabs=false,
    columns=fullflexible,
    keepspaces=true,
    numbers=none,
}

\hypertarget{summary}{%
\section{Summary}\label{summary}}

\href{https://github.com/duncaneddy/brahe}{\texttt{brahe}} is a modern
astrodynamics dynamics library for research and engineering
applications. The representation and prediction of satellite motion is
the fundamental problem of astrodynamics. The motion of celestial bodies
has been studied for centuries with initial equations of motion dating
back to Kepler (1619) and Newton (1687). Current research and
applications in space situational awareness, satellite task planning,
and space mission operations require accurate and efficient numerical
tools to perform coordinate transformations, model perturbations, and
propagate orbits. \texttt{brahe} incorporates the latest conventions and
models for time systems and reference frame transformations from the
International Astronomical Union (IAU) (Hohenkerk, 2017) and
International Earth Rotation and Reference Systems Service (IERS) (Petit
\& Luzum, 2010). It implements force models for Earth-orbiting
satellites including atmospheric drag, solar radiation pressure, and
third-body perturbations from the Sun and Moon (Montenbruck \& Gill,
2000; D. A. Vallado, 2001). It also provides standard orbit propagation
algorithms, including the Simplified General Perturbations (SGP) Model
(D. Vallado et al., 2006). Finally, it implements recent algorithms for
fast, parallelized computation of ground station and imaging-target
visibility (Eddy \& Kochenderfer, 2021), a foundational problem in
satellite scheduling and mission planning.

With \texttt{brahe}, predicing upcoming satellite passes over ground
stations or imaging targets can be accomplished in seconds and three
lines of code.

\begin{lstlisting}[language=Python]
import brahe as bh
bh.initialize_eop()
passes = bh.location_accesses(
    bh.PointLocation(-122.4194, 37.7749, 0.0),  # San Francisco
    bh.celestrak.get_tle_by_id_as_propagator(25544, 60.0, "active"),  # ISS
    bh.Epoch.now(),
    bh.Epoch.now() + 24 * 3600.0,  # Next 24 hours
    bh.ElevationConstraint(min_elevation_deg=10.0)
)
\end{lstlisting}

\texttt{brahe} allows users to quickly access Two-Line Element (TLE)
data from Celestrak (Kelso, T. S., 2025) and propagate orbits using the
SGP4 dynamics model. This can be used to perform space situational
awareness tasks such as predicting the orbits of all Starlink satellites
over the next 24 hours.

\begin{lstlisting}[language=Python]
import brahe as bh
bh.initialize_eop()
starlink = bh.datasets.celestrak.get_tles_as_propagators("starlink", 60.0)
bh.par_propagate_to(starlink, bh.Epoch.now() + 86400.0) # Predict next 24 hours
\end{lstlisting}

The above routine can propagate orbits for all \textasciitilde9000
Starlink satellites in approximately 1 minute 30 seconds on an M1 Max
MacBook Pro with 10 cores and 64 GB RAM. Finally, the package provides
direct, easy-to-use functions for low-level astrodynamics routines such
as Keplerian to Cartesian state conversions and reference frame
transformations.

\begin{lstlisting}[language=Python]
import brahe as bh
import numpy as np

# Initialize Earth Orientation Parameter data
bh.initialize_eop()

# Define orbital elements
a = bh.constants.R_EARTH + 700e3  # Semi-major axis in meters (700 km altitude)
e = 0.001                         # Eccentricity
i = 98.7                          # Inclination in radians
raan = 15.0                       # Right Ascension of Ascending Node in radians
arg_periapsis = 30.0              # Argument of Periapsis in radians
mean_anomaly = 45.0               # Mean Anomaly
state_kep = np.array([a, e, i, raan, arg_periapsis, mean_anomaly])

# Convert Keplerian state to ECI coordinates
state_eci = bh.state_koe_to_eci(state_kep, bh.AngleFormat.DEGREES)

# Define a time epoch
epoch = bh.Epoch(2024, 6, 1, 12, 0, 0.0, time_system=bh.TimeSystem.UTC)

# Convert ECI coordinates to ECEF coordinates at the given epoch
state_ecef = bh.state_eci_to_ecef(epoch, state_eci)

# Convert back from ECEF to ECI coordinates
state_eci_2 = bh.state_ecef_to_eci(epoch, state_ecef)

# Convert back from ECI to Keplerian elements
state_kep_2 = bh.state_eci_to_koe(state_eci_2, bh.AngleFormat.DEGREES)
\end{lstlisting}

Another example application of \texttt{brahe} is predicting and
visualizing GPS satellite orbits. The package provides built-in
functions for generating 2D and 3D visualizations of satellite
constellations using Plotly (Plotly Technologies Inc., 2015) and
matplotlib (Hunter, 2007).

\begin{figure}
\centering
\includegraphics{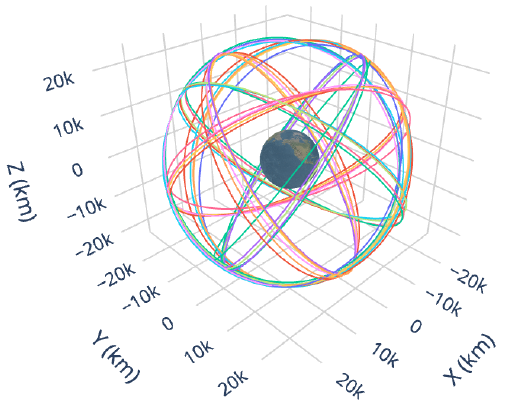}
\caption{Visualization of all GPS Satellite Orbits}
\end{figure}

\hypertarget{statement-of-need}{%
\section{Statement of Need}\label{statement-of-need}}

While the core algorithms for predicting and modeling satellite motion
have been known for decades, there is a lack of modern, open-source
software that implements these algorithms in a way that is accessible to
researchers and engineers. Generally, existing astrodynamics software
packages have one or more barriers to entry for individuals and
organizations looking to develop astrodynamics applications, and often
leads to duplicated and redundant effort as researchers and engineers
are forced to re-implement foundational algorithms.

Flagship commercial astrodynamics software like Systems Tool Kit (STK)
(Analytic Graphics, 2023) and FreeFlyer (a.i. Solutions, Inc., 2025) are
individually licensed and closed-source. The licensing costs can be
prohibitive for researchers, individuals, small organizations, and
start-ups. Even for larger organizations, the per-node licensing cost
can make large-scale deployment prohibitive. The closed-source nature of
these packages makes it difficult to understand and verify the exact
algorithms and model implementations, which is critical for high-stakes
applications like space mission operations (Mars Climate Orbiter Mishap
Investigation Board, 1999). Major open-source projects like Orekit
(Maisonobe et al., 2010) and GMAT (Hughes et al., 2014) provide
extensive functionality, but are large codebases with steep learning
curves, making quick-adoption and integration into projects difficult.
Furthermore, Orekit is implemented in Java, which can be a barrier to
adoption in the current scientific ecosystem with users who are more
familiar with Python. GMAT uses a domain-specific scripting language and
has limited documentation and examples, making it difficult for new
users to get started. Libraries such as poliastro (Cano Rodriguez \&
Martínez Garrido, 2022) and Open Space Toolkit (OSTk) (Open Space
Collective, 2025) provides Python interfaces, but their object-oriented
architecture adds layers of abstraction that can make it difficult to
adapt them to problems that outside their predefined modeling
frameworks. Additionally, poliastro is no longer actively maintained and
OSTk only supports Linux environments and requires a specialized Docker
environment to run. Other academic tools like Basilisk (Kenneally et
al., 2020), provide high-fidelity modeling capabilities for full
spacecraft guidance, navigation, and control (GNC) simulations, but are
not directly distributed through standard package managers like PyPI and
must be compiled from source to be used. Finally, these works often have
limited documentation and usage examples, making it difficult for new
users to get started.

\texttt{brahe} seeks to address these challenges by providing a modern,
open-source astrodynamics library following design principles of the
\emph{Zen of Python} (Peters, 2004). The core functionality is
implemented in Rust for performance and safety, with Python bindings for
ease-of-use and integration with the scientific Python ecosystem.
\texttt{brahe} is provided under an MIT License to encourage adoption
and facilitate integration and extensibility. To further promote
adoption and aid user learning, the library is extensively documented
following the Diátaxis framework (Procida, 2024)\textemdash every Rust
and Python function documented with types and usage examples, there is a
user guide that explains the major concepts of the library, and set of
longer-form examples demonstrating how to accomplish common tasks. To
maintain high code quality, the library has a comprehensive test suite
for both Rust and Python. Additionally, all code samples in the
documentation are automatically tested to ensure they remain functional,
and that the documentation accurately reflects the library's
capabilities.

\texttt{brahe} has already been used in a number of scientific
publications (Eddy et al., 2025; Kim et al., 2025). It has also been
used by aerospace companies such as Northwood Space, Xona Space (Reid et
al., 2020), and Kongsberg Satellite Services for mission analysis and
planning. The Earth Observation satellite imaging prediction and task
planning algorithms have been used by Capella Space and demonstrated
on-orbit with their synthetic aperture radar (SAR) constellation
(Stringham et al., 2019).

\hypertarget{acknowledgments}{%
\section{Acknowledgments}\label{acknowledgments}}

We also want to acknowledge Shaurya Luthra, Adrien Perkins, and Arthur
Kvalheim Merlin for supporting the adoption of the project in their
organizations and providing valuable feedback. Finally, we would like to
thank the Stanford Institute for Human-Centered AI for funding in part
this work.

\hypertarget{references}{%
\section*{References}\label{references}}
\addcontentsline{toc}{section}{References}

\hypertarget{refs}{}
\begin{CSLReferences}{1}{0}
\leavevmode\hypertarget{ref-freeflyer}{}%
a.i. Solutions, Inc. (2025). \emph{{FreeFlyer: Spacecraft Mission
Analysis and Operations Software}} (Version 7.10) {[}Computer
software{]}. \url{https://www.ai-solutions.com/freeflyer}

\leavevmode\hypertarget{ref-stk}{}%
Analytic Graphics, Inc. (AGI). (2023). \emph{{Systems Tool Kit (STK)}}
(Version 12.7) {[}Computer software{]}.
\url{https://www.agi.com/products/stk}

\leavevmode\hypertarget{ref-rodriguezPoliastro2022}{}%
Cano Rodriguez, J. L., \& Martínez Garrido, J. (2022).
\emph{{poliastro}} (Version v0.17.0) {[}Computer software{]}.
\url{https://github.com/poliastro/poliastro/}

\leavevmode\hypertarget{ref-eddyOptimal2024}{}%
Eddy, D., Ho, M., \& Kochenderfer, M. J. (2025). {Optimal Ground Station
Selection for Low-Earth Orbiting Satellites}. \emph{IEEE Aerospace
Conference}. \url{https://arxiv.org/abs/2410.16282}

\leavevmode\hypertarget{ref-eddy2021maximum}{}%
Eddy, D., \& Kochenderfer, M. J. (2021). {A Maximum Independent Set
Method for Scheduling Earth-Observing Satellite Constellations}.
\emph{Journal of Spacecraft and Rockets}, \emph{58}(5), 1416--1429.

\leavevmode\hypertarget{ref-hohenkerk2017iau}{}%
Hohenkerk, C. (2017). {IAU Standards of Fundamental Astronomy (SOFA):
Time and Date}. In \emph{{The Science of Time 2016: Time in Astronomy \&
Society, Past, Present and Future}}. Springer.

\leavevmode\hypertarget{ref-hughes2014gmat}{}%
Hughes, S. P., Qureshi, R. H., Cooley, S. D., \& Parker, J. J. (2014).
{Verification and Validation of the General Mission Analysis Tool
(GMAT)}. \emph{AIAA/AAS Astrodynamics Specialist Conference}.

\leavevmode\hypertarget{ref-Hunter2007}{}%
Hunter, J. D. (2007). {Matplotlib: A 2D Graphics Environment}.
\emph{Computing in Science \& Engineering}, \emph{9}(3), 90--95.
\url{https://doi.org/10.1109/MCSE.2007.55}

\leavevmode\hypertarget{ref-celestrak}{}%
Kelso, T. S. (2025). \emph{{Celestrak} {Active} {Satellite} {Database}}.
\url{https://celestrak.com/}

\leavevmode\hypertarget{ref-basilisk2020}{}%
Kenneally, P. W., Piggott, S., \& Schaub, H. (2020). {Basilisk: A
Flexible, Scalable and Modular Astrodynamics Simulation Framework}.
\emph{Journal of Aerospace Information Systems}, \emph{17}(9), 496--507.
\url{https://doi.org/10.2514/1.I010762}

\leavevmode\hypertarget{ref-kepler1953epitome}{}%
Kepler, J. (1619). \emph{{Epitome Astronomiae Copernicanae}}.

\leavevmode\hypertarget{ref-kim2025scalable}{}%
Kim, G. R., Eddy, D., Srinivas, V., \& Kochenderfer, M. J. (2025).
{Scalable Ground Station Selection for Large LEO Constellations}.
\emph{arXiv Preprint arXiv:2510.03438}.
\url{https://arxiv.org/abs/2510.03438}

\leavevmode\hypertarget{ref-maisonobe2010orekit}{}%
Maisonobe, L., Pommier, V., \& Parraud, P. (2010). {Orekit: An Open
Source Library for Operational Flight Dynamics Applications}.
\emph{International Conference on Astrodynamics Tools and Techniques}.

\leavevmode\hypertarget{ref-mcoMishap1999}{}%
Mars Climate Orbiter Mishap Investigation Board. (1999). \emph{{Mars
Climate Orbiter Mishap Investigation Board Phase I Report}} {[}Tech.
Report{]}. Jet Propulsion Laboratory / National Aeronautics {and} Space
Administration.
\url{https://llis.nasa.gov/llis_lib/pdf/1009464main1_0641-mr.pdf}

\leavevmode\hypertarget{ref-montenbruckgill2000}{}%
Montenbruck, O., \& Gill, E. (2000). \emph{{Satellite Orbits: Models,
Methods and Applications}}. Springer.

\leavevmode\hypertarget{ref-newton1833philosophiae}{}%
Newton, I. (1687). \emph{{Philosophiae Naturalis Principia
Mathematica}}.

\leavevmode\hypertarget{ref-ostk}{}%
Open Space Collective. (2025). \emph{{Open Space Toolkit}}.
\url{https://github.com/open-space-collective/open-space-toolkit}

\leavevmode\hypertarget{ref-peters2004zen}{}%
Peters, T. (2004). {The Zen of Python}. In \emph{{Pro Python}}.
Springer.

\leavevmode\hypertarget{ref-petit2010iers}{}%
Petit, G., \& Luzum, B. (2010). \emph{{IERS Conventions, Technical Note
36}}.

\leavevmode\hypertarget{ref-plotly}{}%
Plotly Technologies Inc. (2015). \emph{{Collaborative Data Science}}.
Plotly Technologies Inc. \url{https://plot.ly}

\leavevmode\hypertarget{ref-procida_diataxis}{}%
Procida, D. (2024). \emph{{Di{á}taxis: A Systematic Approach to
Technical Documentation Authoring}}. \url{https://diataxis.fr/}.

\leavevmode\hypertarget{ref-reid2020satellite}{}%
Reid, T. G., Chan, B., Goel, A., Gunning, K., Manning, B., Martin, J.,
Neish, A., Perkins, A., \& Tarantino, P. (2020). {Satellite Navigation
for the Age of Autonomy}. \emph{2020 IEEE/ION Position, Location and
Navigation Symposium (PLANS)}.

\leavevmode\hypertarget{ref-stringham2019capella}{}%
Stringham, C., Farquharson, G., Castelletti, D., Quist, E., Riggi, L.,
Eddy, D., \& Soenen, S. (2019). {The Capella X-band SAR Constellation
for Rapid Imaging}. \emph{IEEE International Geoscience and Remote
Sensing Symposium}.

\leavevmode\hypertarget{ref-vallado2001fundamentals}{}%
Vallado, D. A. (2001). \emph{{Fundamentals of Astrodynamics and
Applications}}.

\leavevmode\hypertarget{ref-vallado2006revisiting}{}%
Vallado, D., Crawford, P., Hujsak, R., \& Kelso, T. S. (2006).
{Revisiting Spacetrack Report \#3}. \emph{AIAA/AAS Astrodynamics
Specialist Conference and Exhibit}.

\end{CSLReferences}

\end{document}